\documentclass[aps,prl,twocolumn,longbibliography,nopacs,floats,nofootinbib,superscriptaddress]{revtex4-1}
\usepackage{graphicx}
\usepackage{dcolumn}
\usepackage{bm}
\usepackage{amsmath}
\usepackage{color}
\usepackage{amssymb}
\usepackage{ulem}
\usepackage{xcolor}
\usepackage{subfigure}
\usepackage{multibib}

\definecolor{cardinal}{rgb}{0.6,0,0}
\definecolor{darkgreen}{rgb}{0,0.4,0}
\definecolor{golden}{rgb}{0.92, 0.7, 0}
\definecolor{midnight}{rgb}{0, 0, 0.5}
\definecolor{darkblue}{rgb}{0, 0, 0.7}

\def\he4{$^4$He}
\def\hel3{$^3$He}
\def\Am3{\AA$^{-3}$}
\def\beq{\begin{equation}}
\def\eeq{\end{equation}}

\newcommand{\be}{\begin{equation}}
\newcommand{\ee}{\end{equation}}
\newcommand{\bea}{\begin{eqnarray}}
\newcommand{\eea}{\end{eqnarray}}
\newcommand{\bse}{\begin{subequations}}
\newcommand{\ese}{\end{subequations}}

\newcommand{\bkfa}{Ba\textsubscript{1-x}K\textsubscript{x}Fe\textsubscript{2}As\textsubscript{2}}

\begin{document}

\author{Egor Babaev}
\affiliation{Department of Physics, KTH Royal Institute of Technology, Stockholm SE-10691, Sweden}

\author{Boris Svistunov}
\affiliation{Department of Physics, University of Massachusetts, Amherst, MA 01003, USA}
\affiliation{Wilczek Quantum Center, School of Physics and Astronomy and T. D. Lee Institute, Shanghai Jiao Tong University, Shanghai 200240, China}

\title{Hydrodynamics of Borromean Counterfluids}
\begin{abstract}
 Counterflow superfluidity in a system with $N\geq 3$ components is distinctively different from the $N=2$ case. The key  feature
is the difference between the number ($N$) of elementary vortex excitations and the number ($N-1$) of independent branches of phonon modes, that is, the number of  superfluid modes is larger than the number of ordered phase variables. We formulate a hydrodynamic theory  of this state. We show how all the dynamical and statistical aspects of this (``Borromean") type of ordering are naturally described by effective $N$-component theory featuring compact-gauge invariance.
We also discuss how off-diagonal intercomponent couplings convert the Borromean supercounterfluid into a Borromean insulator, with an emphasis on the properties of a non-trivial state with broken time-reversal symmetry.

\end{abstract}

\maketitle

At the microscopic level,  a standard order parameter, or more generally, a classical field that describes a superfluid, is a composite object.  A single complex-valued field with compact phase $\theta \in (0, \, 2\pi]$ perfectly captures the superfluid phenomenology of $^4$He. Nevertheless, this field emerges out of six fermionic fields: pairs of electronic, protonic, and neutronic fields (as long as nucleons are treated as elementary particles).
In recent decades, much attention has been devoted to phenomena arising in multicomponent systems, which are principally different from single-component ones, with the description requiring multiple complex-valued fields. 

The main focus of the research on multicomponent systems (for a review, see Ref.~\cite{Svistunov2015}) was on the two-component case.
One new phenomenon that arises is the counterflow superfluidity, which can emerge as a result of a phase transition out of two-component superconductor \cite{babaev2002phase,babaev2004superconductor}, as well as out of two-component superfluid near Mott insulating state \cite{kuklov2003counterflow}. The supercounterfluid {\it ground state} 
can take place in a lattice system at a commensurate net filling \cite{kuklov2003counterflow} (see also Conclusions and Discussion section).
The key difference between these systems and ordinary superfluids is that only the relative motion of the two components is dissipationless. (For an early discussion of the $N$-component case, see, e.g., Ref.~\cite{Smiseth2005field}.)

Recent exciting experiments
\cite{Grinenko2021state,shipulin2023calorimetric} reported a discovery of counterflow order in an iron-based material \bkfa. That particular order manifests itself as a new phase---forming prior to the superconducting phase transition---with broken $Z_2$ discrete time-reversal symmetry \cite{Bojesen2013time,Bojesen2014phase}. Microscopically, the breakdown of time-reversal symmetry is associated with the formation of nontrivial phase differences between phases of non-condensed electronic pairs in different bands. The fact that only a discrete symmetry is broken implies that associated counterflow currents have only a short range. The newly discovered phase exhibits multiple unconventional phenomena in transport, thermoelectric, and ultrasound probes.
There is an experimental evidence in favor of existence of three flavors of Cooper pairs in the superconducting state of this compound \cite{Grinenko2017bkfa, Grinenko2020superconductivity}.
Additionally, fractional vortices were observed in the same compound \cite{Iguchi2023} thus establishing its multicomponent nature. Recently, three-component counterflow superfluidity was revealed in first-principles simulations of bosonic lattice model \cite{blomquist2021borromean}.

 Curiously enough, the   experimental and theoretical progress left behind a fundamental question: What is the effective hydrodynamic ground-state description of a Borromean counterfluid?  A simple ``counting" argument immediately tells us that such a description, if at all possible, should be rather special. Given that the net matter flow is arrested, the number of  independent phonon modes has to be equal to $N-1$, suggesting that the number of independent order parameters---defining the number of pairs of canonically conjugate fields in the Hamiltonian formalism---is also $N-1$. On the other hand, the number of elementary vortices, or, equivalently, elementary counterflow persistent-current states equals $N$ (provided $N \geq 3$), as if we had 
$N$ independent order parameters. Furthermore, the very idea of constructing a translation-invariant Hamiltonian---as it should be in the long-wave limit when the microscopic details leading to the suppression of the net matter flow become irrelevant---may seem questionable. Meanwhile an arrest of bulk current is possible in a very different kind of translation-invariant system: local gauge theory. It is nothing but Meissner effect   in the London-Ginzburg-Landau theory of superconductivity \cite{london1948problem,Ginzburg1950}, where the gauge invariance of the theory, while allowing for topological excitation (vortices), leads to Anderson effect \cite{Anderson1963} of elimination of Goldstone mode and supertcurrent in the bulk of the system. However, the system we consider has no local gauge symmetry.

In this Letter, we formulate $N$-component counterflow hydrodynamic Hamiltonian yielding a very transparent picture of the elementary and topological excitations of counterfluids. The central feature of the effective theory is the new kind of gauge invariance that we call compact-gauge invariance. It eliminates (gauges out) the net flow of the components despite translation invariance of the system. At $N \geq 3$,  compact-gauge invariance still preserves all $N$ elementary topological excitations/persistent currents---by which we mean a regime when component  $\alpha$ has a (well-defined) $\pm 2\pi$ relative phase winding with respect to the other $N-1$ components, or equivalently (due to the compact-gauge redundancy) all the $N-1$ components $\beta \neq \alpha$ have a $\mp 2\pi$ relative phase winding with respect to the component $\alpha$. Importantly, the symmetry that this system breaks is not the conventional [U(1)]$^{N-1}$.

{\it Counterflow hydrodynamic Hamiltonian.}   {Counterflow of two components is described as the gradient of their phase differences.
The corresponding free energy for the three-component case
 was established numerically \cite{blomquist2021borromean}.
} Here, we observe that, in the linear limit, the long-wave {\it dynamics} of an $N$-component supercounterfluid  is described by the following bilinear Hamiltonian density
\be
{\cal H} \, =\, {1\over 2} \sum_{\alpha , \beta} \kappa_{\alpha \beta} \, \eta_\alpha \eta_\beta \, +\, 
{1\over 2} \sum_{\alpha < \beta} \Lambda_{\alpha  \beta} (\nabla \theta_\alpha \! - \! \nabla\theta_\beta)^2 \, ,
\label{H_SCF}
\ee
where $(\eta_\alpha, \theta_\alpha)$, $\alpha =1,2,3,\ldots N$, is a pair of canonically conjugate variables for the $\alpha$-component: $\theta_\alpha$ is the superfluid phase
and $\eta_\alpha$ is the deviation of the density of this component from the equilibrium (ground-state) value. Without loss of generality, the matrices $\kappa_{\alpha \beta} $ and $\Lambda_{\alpha  \beta}$ are assumed to be symmetric. We  also require that the quadratic forms in  (\ref{H_SCF}) be nonnegative-definite.  What we explore here are the dynamical implications of promoting the structure of free energy demonstrated numerically  in Ref.~\cite{blomquist2021borromean}---the second term in the right-hand side of (\ref{H_SCF})---to the Hamiltonian by adding the first term containing the densities as the variables canonically conjugate to the phases. We deliberately start with the special (limiting) case of the hydrodynamic Hamiltonian; generalization, however, proves rather straightforward, as we will see later.

Hamiltonian (\ref{H_SCF}) generates the following equations of motion:
\bea
\dot{\theta}_\alpha & =&  - \sum_{\beta}  \kappa_{\alpha \beta} \,  \eta_\beta \, ,
\label{em_1}
\\
\dot{\eta}_\alpha & =&  - \sum_{\beta}   \Lambda_{\alpha \beta} \,   (\Delta \theta_\alpha \! - \! \Delta \theta_\beta)\, .
\label{em_2}
\eea
Summing up equations (\ref{em_2}) for all the components, we obtain a local conservation law,
\be
{\partial \over \partial t} \sum_{\alpha = 1}^N \, \eta_\alpha({\bf r},t) \, =\, 0  \, ,
\label{constr}
\ee
revealing one of the most desired properties of the system: The net local density stays unchanged, even if initially perturbed!
Therefore, one of the normal modes---the one corresponding to Eq.~(\ref{constr})---proves to be pathological. We thus conclude that there are only $N-1$ branches of phonon excitations. In certain microscopic realizations of the counterflow superfluids, weakly perturbing the net density might be fundamentally impossible. In such cases, constraint (\ref{constr}) should be understood as the relation demonstrating the hydrodynamic consistency of the condition
\[
\sum_{\alpha = 1}^N \, \eta_\alpha({\bf r},t) \, \equiv \, 0  \, .
\]
 
Equations (\ref{em_2}) can be interpreted as the componentwise continuity equations
 \be
 \dot{\eta}_\alpha \, +\, \nabla \cdot{\bf j}_\alpha \, =\, 0 \, ,
 \label{cont}
 \ee
 where 
${\bf j}_\alpha  \, = \, \sum_{\beta}  \,  \Lambda_{\alpha \beta} \,   (\nabla \theta_\alpha \! - \! \nabla \theta_\beta)$ 
is the current density of the component $\alpha$. Correspondingly, constraint (\ref{constr}) can be viewed as an immediate implication of the absence of the net current,
\be
\sum_\alpha \,  {\bf j}_\alpha  \, = \, 0 \, .
\label{zero_j_net}
\ee
Condition (\ref{zero_j_net})
is the defining feature behind the term ``counterfluid."

Constraint (\ref{constr})---or, equivalently, the counterflow condition  (\ref{zero_j_net})---has a fundamental origin. It is the Noether's constant of motion corresponding to the gauge symmetry of the problem---the invariance of the Hamiltonian density (\ref{H_SCF}) with respect to the transformation
\be
\forall \alpha : \quad \theta_\alpha ({\bf r}) \, \to \,  \theta_\alpha ({\bf r}) \, +\, \phi ({\bf r}) \, .
\label{gauge}
\ee
 There are important differences compared to the standard U(1) gauge theories,
  rendering the class of legitimate gauge transformations special---with certain extra restrictions and certain extra (and most relevant) freedom. In the absence of U(1) gauge field, the field $\phi ({\bf r})$ cannot depend on time, as is immediately clear from Eq.~(\ref{em_1}). Hence, the time derivatives of the phases are gauge-invariant quantities.
In a fundamental contrast with the standard gauge transformation of the phases of complex-valued matter fields in U(1) gauge theories, where the field $ \phi ({\bf r})$ is supposed to be smooth and single-valued, it is essential that $ \phi ({\bf r})$ in (\ref{gauge}) is understood as the {\it compact field of phase} allowed to have topological defects (vortices), as well as global phase windings (in toroidal systems).
Furthermore, within our  effective long-wave descriptions, these defects (that can be thought of as vortices with same windings in all the components) cost no energy. It is thus fair to say that in the long-wave limit,
the only  physically relevant (gauge-invariant) quantities are the {\it relative} phases
\be
\Phi_{\alpha \beta} \, =\, \theta_\alpha - \theta_\beta \, .
\label{relative}
\ee

{

{\it Modular arithmetic of topological charges.}  Compact-gauge-redundant description is very convenient for classification, evaluation, and comparison of topological charges of  elementary and composite topological defects and supercurrent states. Formally ascribing an individual integer topological charge (phase winding number) to each of the $N$ components, we characterize the charge of a system of topological defects or/and supercurrent states by corresponding string of integers: $(m_1,\, m_2,\, \ldots , \, m_N)$ (cf.~\cite{Smiseth2005field}). The ambiguity of this description is fully exhausted by the  compact-gauge redundancy thus leading to the {\it modular arithmetic of topological charges}. The  addition/equality of the strings of topological charges is performed modulo $(1,\, 1,\, \ldots , \, 1)$. Example 1: $(1,0,0)+(0,1,0)=(1,1,0) \equiv (0,0,-1)$.  
That is, a superposition of $N-1$ elementary defects of the same sense is topologically equivalent to the $N$-th elementary defect of the {\it opposite} winding.
Example 2: $(1,0,0)+(0,1,0)+(0,0,1)=(1,1,1) \equiv (0,0,0)$, the superposition of $N$ elementary defects of the same sense is topologically neutral.

It is important to emphasize that modular arithmetic of topological charges does not imply that the number of elementary vortices reduces to $N-1$. An elementary vortex is a distinct soliton-type object minimizing the energy in corresponding topological sector and therefore being  energetically protected from a decay into a superposition of other excitations. In a system with $N\geq 3$ components, we have exactly $N$ such excitations. It is very instructive to note that this crucial fact has a direct   statistical implication in two dimensions, where the supercounterfluid-to-normal transition  is of  Berezinkskii-Kosterltz-Thouless  type \cite{tobepubl}.

{\it   Borromean insulators: Counterfluid with symmetry-breaking off-diagonal intercomponent couplings.} Consider a generalization of the Hamiltonian (\ref{H_SCF}) that includes intercomponent Josephson couplings:
\be
{\cal H} \, \to \, {\cal H}\, - \, {1\over 2} \sum_{\alpha, \beta} {\cal Q}_{\alpha  \beta} (\theta_\alpha \! - \! \theta_\beta) \, ,
\label{H_CF_J}
\ee
where ${\cal Q}_{\alpha  \beta} (\theta) = {\cal Q}_{ \beta \alpha } (\theta) = {\cal Q}_{ \alpha \beta } (-\theta)$ are $2\pi$-periodic even real functions of the argument $\theta$. Speaking generally, these functions also depend on the density deviations, 
 which, under certain conditions, leads to various effects such as mode mixing
\cite{Carlstrom2011length}. However, the details of this dependence are not relevant to our discussion; note also the analysis of the general case presented below. 

With the new Hamiltonian, the equations for $\dot{\eta}$---conveniently written in the form of generalized continuity equations (\ref{cont})---become
 \be
 \dot{\eta}_\alpha \, +\, \nabla \cdot{\bf j}_\alpha \, =\, {\cal J}_\alpha \, ,
 \label{cont2}
 \ee
 with the intercomponent Josephson currents
  \be
{\cal J}_\alpha = \sum_{\beta} {\cal W}_{\alpha  \beta} (\theta_\alpha \! - \! \theta_\beta)\, , \quad 
{\cal W}_{\alpha  \beta} (\theta) = {\partial {\cal Q}_{\alpha  \beta} (\theta)\over \partial \theta}  \, .
 \label{J_alpha}
 \ee
 As is seen from (\ref{J_alpha}), these currents sum up to zero,
   \be
\sum_\alpha \, {\cal J}_\alpha \, =\, 0 \, ,
 \label{J_net_zero}
 \ee
which, in combination with (\ref{zero_j_net}), leads to the local conservation law (\ref{constr}). The fundamental origin of the conservation law is still the same---invariance of the Hamiltonian with respect to the gauge transformation (\ref{gauge}).

In view of the constraint (\ref{constr}), we can still refer to the model (\ref{H_CF_J}) as a counterfluid, especially when talking of the normal modes: These   involve  counterflow motion of the components without changing the net density.

{\it  Borromean insulators and metals with broken time-reversal symmetry. {
Permutational ordering.}} 
The equations of motions generated by Hamiltonians (\ref{H_SCF}) and (\ref{H_CF_J}) feature time-reversal symmetry,
$t \, \to\, -t \, , \qquad \forall \alpha: ~~ \theta_\alpha \, \to\, -\theta_\alpha \, ,$ 
which the ground state of the Hamiltonian (\ref{H_CF_J}) breaks  {
whenever the form of the functions ${\cal Q}_{\alpha  \beta}(\theta)$ is such that energy minimum takes place at $0< \Phi_{\alpha \beta} < \pi$ for two or more components. Here, the time-reversal symmetry breaking is
a particular case of more general order.
Suppose we have an $N$-component Borromean insulator the Hamiltonian of which features (i)  exact permutational symmetry between all the components and (ii) a ground states with $\Phi_{21}=\Phi_{32}= \ldots =\Phi_{1 N}=2\pi/N$. By the symmetry between the components, such a ground state is $(N\!-\!1)!$-fold degenerate. At $N>3$ with higher degeneracy, the broken time-reversal symmetry is only a part  of this  {\it spontaneously broken permutational symmetry} \cite{Weston2013}. 
 
}
{
When the time-reversal  (or permutaion) symmetry is broken,
the system supports more than one  kind of domain walls. Even when the broken symmetry is $Z_2$ at $N=3$, there are still more than two domain walls because there are more than two ways to interchange phases (cf.~\cite{Garaud2013}).}
Domain walls imply gradients of the relative phases and hence various patterns of
 persistent counter-currents. As was mentioned above, dissipationless DC currents are not possible.

 {  Several remarks are in order here. {Different realizations of Borromean states involve different mechanisms of the arrest of the net superflow. When the arrest is due to Mottness \cite{kuklov2003counterflow}, the system is a ground-state insulator. However, at the macroscopic scale, the situation is similar to the case when the arrest of the net bulk supercurrent is enforced by the gauge-field coupling \cite{babaev2002phase,babaev2004superconductor,Grinenko2021state}. In the later case, when the Borromean state occurs above superconducting phase transition,
 it allows dissipative single-electron-based current, i.e., it is a Borromean metal. However, the ground-state Borromean insulator also crosses over---with increasing temperatures---to the state of Borromean metal.
 
Speaking about metallic versus insulating states, we note that a conventional superconductor can be characterized as a peculiar topological-insulator-like state. Indeed,
due to the Anderson (Meissner) effect, the superconductor has no currents in its bulk, while allowing Meissner currents at its surface. 
{ In the multicomponent BTRS (broken time-reversal symmetry) system \bkfa, persistent counterflow currents  are observed 
both in the Borromean metal and below the superconducting transition in 
  ~\cite{Grinenko2020superconductivity,Grinenko2021state}.
In a superconducting state, defects act as boundaries and induce both the counterflow and Meissner screening currents \cite{garaud2014domain}.
Below the superconducting phase transition, the currents are not purely Borromean, due to conventional local gauge invariance leading to creation of Meissner screening currents.
   Importantly, there is a genuine phase transition between a BTRS superconducting state with persistent currents (which is an insulator in the bulk---in the Anderson's sense) and the state which has persistent Borromean currents and the resistive DC transport  discussed in \cite{Grinenko2021state}. }
   In one case, the system has conventional gauge invariance, in the other case, the system features compact gauge invariance. 
}

{\it General form of the counterflow Hamiltonian.} Consider Hamiltonian density given by the following function

\be
{\cal H} \, \equiv \, {\cal H}   (\{ \eta \}, \{ \theta \}, \{ {\bf v} \}) \, ,
\label{H_SCF_gen}
\ee
where  the arguments $\{ \eta \} \equiv (\eta_1, \eta_2, \ldots \eta_N)$ and $\{ \theta \} \equiv (\theta_1, \theta_2, \ldots \theta_N)$  are placeholders for corresponding fields, while $ \{ {\bf v} \} \equiv ({\bf v}_1, {\bf v}_2, \ldots {\bf v}_N)$ are placeholders for the fields of phase gradients: ${\bf v}_\alpha \to \nabla \theta_\alpha$. 

For the theory to be invariant with respect to the gauge transformation (\ref{gauge}), we require that  ${\cal H}$ satisfy the following two conditions
\be
\sum_\alpha \, {\partial {\cal H}\over \partial \theta_\alpha} \, =\, 0\, , \qquad 
\sum_\alpha \, {\partial {\cal H}\over \partial {\bf v}_\alpha} \, =\, 0\, .
\label{gen_rel}
\ee
The equations of motion for the fields $\eta_\alpha$ come in the form of generalized continuity equations (\ref{cont2}) with 
\be
{\bf j}_\alpha  \, = \, {\partial {\cal H}\over \partial {\bf v}_\alpha} \, , \qquad 
{\cal J}_\alpha \, = \, {\partial {\cal H}\over \partial \theta_\alpha}
\qquad  (\{ {\bf v} \}\to \{ \nabla \theta \} ) \, .
\label{j_gen}
\ee
Conditions (\ref{gen_rel}) then lead to the relations (\ref{zero_j_net}) and (\ref{J_net_zero}), thus implying  (\ref{constr}).

{\it Conclusions and discussion.} We established  effective hydrodynamic Hamiltonian for multicomponent systems featuring ground states with  Borromean counterflow ordering. The basic version of the Hamiltonian, Eq.~(\ref{H_SCF}), describes
the normal modes (phonons) and topological excitation---vortices and persistent-current states---of counterflow superfluid. The extended version, Eq.~(\ref{H_CF_J}), describes the system with the off-diagonal coupling between components, in which case the ground state of the system does not break continuous symmetry but, in certain cases, can feature Borromean-type breaking of the permutation symmetry, which in the simplest case is equivalent to breaking time-reversal symmetry. The   general form of the counterflow hydrodynamic Hamiltonian is given by Eqs.~(\ref{H_SCF_gen})--(\ref{gen_rel}). 
The key part of our theory is played by  the concept of compact-gauge symmetry, Eq.~(\ref{gauge}), enforcing the counterflow character of the motion of the components, irrespective of the type of symmetry breaking (if any). In this connection, it is instructive to make a comparison with the Anderson effect of eliminating a Goldstone mode by opening a gap \cite{Anderson1963}. In our case, the effect is different and, in a way, more consequential in the sense that the mode is completely eliminated rather than rendered gapped.

As an effective long-wave model, our classical-field Hamiltonian is  insensitive to the microscopic origin of the counterflow ground states. However, in order to complete the physical picture, we refer to one microscopic setup, resulting in such a description in the long-wave limit.  As it was argued in Ref.~\cite{kuklov2003counterflow} (see also \cite{Svistunov2015}), the counterflow ground state naturally emerges in a multicomponent lattice bosonic or fermionic system at a commensurate filling, when the interaction between particles is strong enough to create the Mott gap for the net particle transport. While suppressing the net particle flow, the Mott gap still allows for the counterflow transport via the super-exchange mechanism.   A minimal model capturing all the physics we discussed in this work is a three-component fermionic model with [U(1)]$^3$---as opposed to SU(3)---symmetry. In order to produce BTRS ground state, the [U(1)]$^3$ symmetry is further reduced  by introducing Josephson intercomponent couplings.  A distinctively instructive aspect of such a model is that the (super-)counterflow regime cannot be consistently  described in terms of two-particle pairing(s). ``Borromean" ordering of fermions represents the generalization of superfluidity and superconductivity beyond the current paradigms based on pairing
and breaking local or global U(1) symmetry. Namely for more than two components, this class of states features dissipatiponless modes with
manifestly absent pair formation (i.e. species of counterflowing fermions are not independently conserved).

In the experiment of Ref.~\cite{Grinenko2021state}, the Borromean metallic state emerges on top of BTRS superconducting state, from which it is separated by a finite-temperature phase transition. 
 However, here we stress that, in general, one cannot describe the composite orders considered here or quartic metal in terms of a  small fluctuation-induced ``vestigial"  order.
In a broader context, the state that we call  Borromean metal naturally sets on top of the ground state of BTRS Borromean insulator as a result of a finite-temperature insulator-to-metal crossover, and is not necessarily the result of partial restoration of symmetry via thermal phase transition.

\section{Acknowledgements} We thank Anatoly Kuklov and Nikolay Prokof'ev for the discussion of our results.
This work is supported by the Swedish Research Council Grants 2016-06122 and 2022-04763. BS thanks the Department of Physics of KTH for hospitality during his visit resulted in this work; he also acknowledges support from the National Science Foundation  under Grants  DMR-2032077 and  DMR-2335904.
%


\end{document}